\documentclass[runningheads]{llncs}
\usepackage[T1]{fontenc}
\usepackage{amsmath}
\usepackage{amssymb}
\usepackage{graphicx}
\usepackage{verbatim} 
\usepackage{tikz}
\usetikzlibrary{shapes}
\begin{document}
\title{An Executable Specification of Oncology
  Dose-Escalation Protocols with Prolog}
\titlerunning{Executable Specification of Dose-Escalation Protocols with Prolog}
%

\author{
  David C. Norris\inst{1}
  \and
  Markus Triska\inst{2}
}
\authorrunning{D. Norris and M. Triska}
\institute{Precision Methodologies, LLC, Wayland MA 01778, USA\\
\email{david@precisionmethods.guru}\\
\url{https://precisionmethods.guru} \and
\email{triska@metalevel.at}\\
\url{https://www.metalevel.at}}

\maketitle              
\bibliographystyle{splncs04}

\begin{abstract}
We present, as a pure Prolog program, the first executable
specification of the \mbox{3 + 3} dose-escalation protocol commonly
used in early-phase oncology drug development.  In this program, the
imperative operations of the protocol emerge as consequences of
clinically meaningful anticipatory-regret scenarios that are declared
as CLP($\mathbb{Z}$) constraints.  This `regret-constrained' (RC)
specification yields a robust formulation which can be used to prove
clinically meaningful safety and liveness properties of the protocol
before incorporating it into a trial, and then as an on-line decision
support system while the trial is underway.  Our RC specification also
readily accommodates certain pragmatic modifications to trial
enrollment which severely strain traditionally imperative
formulations.  The features of modern Prolog systems let us describe
the \mbox{3 + 3} protocol with a short and general program that has
desirable algebraic properties and can therefore be used, tested and
reasoned about in several different ways.

\keywords{System safety \and Formal system verification \and Clinical
  trials \and Dose finding.}
\end{abstract}
\section{Introduction}

Clinical development of a new cancer drug for human use typically
follows a 3-phase process. Phase~1 trials, which concern us here, try
a drug for the first time in humans, exploring relations between {\em
  dosing} and {\em toxicity} which preclinical animal experimentation
cannot adequately characterize.  Subsequent phase~2 and 3 trials
address {\em efficacy}, and are premised on dosing recommendations
yielded by phase~1.  Because cancer drugs' toxicity usually becomes
evident on a shorter time-scale than their efficacy, phase~1 cancer
trials often adopt a {\em maximum tolerated dose} (MTD) heuristic
which calibrates dosing to produce mild-to-moderate toxicity.  While
it has long been recognized that MTD varies from one patient to
another \cite{edler_statistical_1990}, it remains an almost universal
practice in phase 1 oncology trials to define `tolerability' in {\em
  population} terms, according to what proportion of patients
experience a severe, `dose-limiting' toxicity (DLT).

\section{Prolog prerequisites and compatibility considerations}

Our key contribution, described in the following sections, is the
specification of a dose-escalation protocol in the form of an
executable Prolog~program. We have chosen Prolog as implementation
language for several reasons: First, Prolog is highly suited for this
application, allowing us to achieve a general and efficient program
with a small code-base. Second, Prolog remains the best-known
logic programming language, continuing to be
taught at universities around the world. This makes our
specification immediately accessible to a
substantial number of programmers. Third, a number of
different Prolog implementations exist, and yet are mutually
compatible to varying degrees thanks to the existence of an
ISO~standard that defines their common syntactic and semantic
features.

Standards are of great importance in the medical sector and play a
significant role in procurement decisions, resolution of legal
disputes, warranty questions, and the preparation of
teaching~material. It is to be expected that the use of an
ISO-standardized programming language will enable the broadest
possible adoption of our approach in such a safety-critical
application~area. For these reasons, we are using Scryer~Prolog for
our application. Scryer Prolog is a modern Prolog system written in
Rust that aims for strict conformance to the Prolog ISO~standard and
satisfies all syntactic conformity tests given in
\textit{https://www.complang.tuwien.ac.at/ulrich/iso-prolog/conformity\_testing}.
It is freely available from: \textit{https://www.scryer.pl}. The
queries we show are executed with version~0.9.3.

A moderate knowledge of Prolog is required to fully understand the
code presented here. We are using only basic Prolog predicates, and
features from the following additional libraries that are available in
Scryer Prolog and also for several other Prolog systems:

\begin{verbatim}
:- use_module(library(lists)).
\end{verbatim}
 Commonly known relations over lists, such as \texttt{length/2} and
 \texttt{maplist/N}.
\begin{verbatim}
:- use_module(library(clpz)).
\end{verbatim}
 Declarative integer arithmetic, with a \textit{monotonic} execution
 mode requiring that all logic~variables standing for concrete integers
 be wrapped with the dedicated functor~\texttt{(\#)/1}.  We use
 CLP($\mathbb{Z}$) constraints such as \texttt{(\#=)/2} and
 \texttt{(\#>)/2} to denote the respective relations between integer
 expressions, allowing for very general definitions.
\begin{verbatim}
:- use_module(library(reif)).
\end{verbatim}
This library provides the meta-predicates \texttt{if\_/3} and
\texttt{tfilter/3} as described
in~\cite{DBLP:journals/corr/NeumerkelK16}.
\begin{verbatim}
:- use_module(library(dcgs)).
\end{verbatim}
 Definite clause grammars (DCGs) which are available in virtually all
 Prolog systems.  We are using this built-in grammar mechanism to
 describe \textit{sequences} of events.  The nonterminal
 \texttt{... //0} means \textit{any sequence}.
\begin{verbatim}
:- use_module(library(error)).
\end{verbatim}
This library provides sound type tests.  If a term is not sufficiently
instantiated to decide its type, an instantiation error is raised to
prevent incorrect failure in cases where a more specific instantiation
would yield a solution.
\begin{verbatim}
:- use_module(library(si)).
\end{verbatim}
``si'' stands for \textit{sufficiently instantiated}, and also for
\textit{sound inference}.  This library provides additional sound type
tests.
\begin{verbatim}
:- use_module(library(lambda)).
\end{verbatim}
 We use this higher-order construct to conveniently \textit{project
   away} uninteresting answer~substitutions in toplevel interactions,
 and also to define anonymous auxiliary predicates that do not warrant
 a standalone definition or dedicated name.
\begin{verbatim}
:- initialization(assertz(clpz:monotonic)).
\end{verbatim}
  We run our application in the monotonic execution mode to ensure
  that all admissible cases are taken into account, and consequently
  use for example the term~\texttt{\#X} to denote a logic
  variable~\texttt{X} that stands for a concrete integer.

We provide a collection of texts that explain these language features
in more detail at \textit{https://www.metalevel.at/prolog}.
These language features are explained in more detail at
\textit{https://www.scryer.pl} and the resources it links to.

The Prolog code we present in the following sections is designed to
run with all conforming Prolog~systems where the above provisions are
present.

\section{Dose-escalation trials}\label{sec:mechanics}

Participants in phase~1 oncology studies are typically patients with
cancer for whom standard treatments have stopped working or do not
exist.  These patients enroll to pursue an experimental treatment with
{\em therapeutic intent} \cite{weber_reaffirming_2016} in the face of
unknown efficacy and great uncertainty about how toxicities may
manifest and at what doses.

The required therapeutic balance between tolerability and efficacy is
typically sought in {\em dose-escalation} trials that enroll patients
serially into an escalating sequence of pre-specified doses.  Small
{\em cohorts} of 1--3 patients are enrolled together, all assigned to
the same dose level.  Although we will demonstrate a pragmatically
important generalization to variable cohort sizes, initially we specify
uniform cohorts of size 3.
\begin{verbatim}
allowed_cohort_sizes([3]).
\end{verbatim}
Whereas a complete trial protocol will specify complex clinical
grading systems for assessing a broad array of distinct toxicities,
for dose-escalation purposes each patient's experienced
toxicity gets tallied in binary terms, according to whether the
toxicity is severe enough to be deemed a `dose-limiting toxicity'
(DLT).  Eliding the time which must elapse from dosing to toxicity
assessment, at any given moment each dose will be characterized by a
cumulative
{\em toxicity tally} {\tt T/N} denoting that a DLT has occurred in \verb"T" out of the \verb"N"
participants enrolled at that dose:
\begin{verbatim}
valid_tally(T/N) :- valid_denom(N), 0 #=< #T, #T #=< N.
\end{verbatim}
Rapid progress being imperative in cancer drug development, phase~1
trials are typically designed to yield a recommended phase~2
dose
without enrolling a large number of patients.  Thus, {\em for phase~1
  purposes}, the maximum enrollment deemed necessary to characterize
any given dose may be quite modest:
\begin{verbatim}
valid_denom(N) :- N in 0..6.
\end{verbatim}
If a dose has current tally {\tt T0/N0}, then enrollment of a new
cohort at this same dose will yield a new tally {\tt T/N} in the
expected way:
\begin{verbatim}
enroll(T0/N0, T/N) :-
    allowed_cohort_sizes(Cs),
    member(Nnew, Cs),
    #N #= #N0 + #Nnew,
    valid_denom(N),
    Tnew in 0..Nnew, indomain(Tnew),
    #T #= #T0 + #Tnew.
\end{verbatim}
The dose-escalation {\em state} at any moment consists of
accumulated toxicity tallies at each of an ascending sequence of up to
8 doses (a pragmatically relevant maximum
rarely exceeded by actual dose-escalation trials),
together with an index into this sequence identifying a
`current dose'.  We represent this by a pair of tally-lists, the
left-hand \verb"Ls" listing lower doses in {\em descending} dose order with
the current dose at its head, and the right-hand \verb"Hs" listing
higher doses in {\em ascending} order with the next-higher dose at its head.
\begin{verbatim}
valid_state(Ls - Hs) :-
    append(Ls, Hs, Ds),
    length(Ds, ND), ND in 1..8,
    maplist(valid_tally, Ds).

state_tallies(Ls-Hs, Qs) :-
    valid_state(Ls-Hs),
    reverse(Ls, Js),
    append(Js, Hs, Qs).
\end{verbatim}
As tallies accumulate at the various dose levels, the `current dose
level' (into which the next cohort is to be enrolled) may be updated
by one of 3 possible dose-escalation {\em decisions}: {\tt esc}alate,
{\tt sta}y and {\tt d}e-{\tt es}calate:
\begin{verbatim}
state0_decision_state(Ls - [H0|Hs], esc, [H|Ls] - Hs) :-
    valid_state(Ls - [H0|Hs]),
    enroll(H0, H).
state0_decision_state([L0|Ls] - Hs, sta, [L|Ls] - Hs) :- 
    valid_state([L0|Ls] - Hs),
    enroll(L0, L).
state0_decision_state([L,D0|Ls] - Hs, des, [D|Ls] - [L|Hs]) :- 
    valid_state([L,D0|Ls] - Hs),
    enroll(D0, D).
\end{verbatim}

The very {\em data structures} in which we have declared the basic
template of a dose-escalation protocol already limit which
dose-escalation decisions are {\em feasible} at any point in the
trial.  One cannot, for instance, escalate above the highest dose; nor
can one de-escalate from the lowest:
\begin{verbatim}
state0_decision_infeasible_t(_-[], esc, true).
state0_decision_infeasible_t([_]-_, des, true).
\end{verbatim}
We encode maximum dose-wise enrollment also
as a matter of {\em feasibility}:
\begin{verbatim}
state0_decision_infeasible_t(_-[_/N|_], esc, true)   :- #N #>= 6.
state0_decision_infeasible_t([_/N|_]-_, sta, true)   :- #N #>= 6.
state0_decision_infeasible_t([_,_/N|_]-_, des, true) :- #N #>= 6.
\end{verbatim}
The \verb"false" clauses of \verb"state0_decision_infeasible_t/3" are
obtained by demonstrating that at least one valid subsequent state {\em does} exist:
\begin{verbatim}
state0_decision_infeasible_t(S0, E, false) :-
    member(E, [esc,sta,des]),
    Goal = exists_subsequent_state(S0, E),
    once((term_si(Goal), Goal)).

exists_subsequent_state(S0, E) :-
    state0_decision_state(S0, E, _).
\end{verbatim}
The existential quantification performed here uses the
(generally, impure) predicate \verb"once/1" to avoid redundant
solutions for greater efficiency, yet it does so in a safe~way:
\verb"term_si/1" ensures that \verb"once/1" is only called when
its~argument is ground and can therefore yield at most one solution.
Importantly, \verb"term_si/1" yields an \textit{instantiation error}
if its argument is not ground, differing crucially from the
standard predicate \verb"ground/1", which can fail silently even in
cases that admit solutions by further instantiations.  Therefore our
use of \verb"once/1" {\em never incurs silent loss of solutions.}

In addition to the \verb"esc", \verb"sta" and \verb"des" decisions
which permit the trial to continue, there is also the possibility of
{\tt stop}ping with a dose recommendation.  The {\em rules} which
relate observed toxicity tallies to these decisions constitute the
{\em design} of a dose-escalation protocol.

\newpage
\section{The traditional rules of 3 + 3 dose-escalation}

The conventional, and still most prevalent, dose-escalation design is
the \mbox{3 + 3}.  For an account of this design {\em in practice}, we
are indebted to \cite{korn_comparison_1994}, quoted at length in
Figure~\ref{korn_protocol}. A similar description from
\cite{skolnik_shortening_2008} is given in
Figure~\ref{skolnik_protocol}.

\begin{figure}
\begin{quote}
{\em ``Although there is no canonical ‘standard method’, we present the
  design that we have most often encountered, which is used almost
  universally with minor variations. First, there is a precise
  [clinical] definition in the protocol of what is considered
  dose-limiting toxicity (DLT); it may differ in different settings. The
  dose levels are fixed in advance, with the first patients treated at
  dose level~1.  Initially, one treats 3 patients at a level. If 0 out
  of 3 patients experiences DLT, one proceeds to the next higher level
  with a cohort of 3 patients. If 1 out of 3 patients experiences DLT,
  treat an additional 3 patients at the same dose level. If 1 out of 6
  patients experiences DLT at a level, the dose escalates for the next
  cohort of patients. If $\ge 2$ out of 6 patients experience DLT, or
  $\ge 2$ out of 3 patients experience DLT in the initial cohort treated
  at a dose level, then one has exceeded the MTD. Some investigations
  will, at this point, declare the previous dose level as the MTD, but a
  more common requirement is to have 6 patients treated at the MTD (if
  it is higher than level 0). To satisfy this requirement, one would
  treat another 3 patients at this previous dose level if there were
  only 3 already treated. The MTD is then defined as the highest dose
  level $(\ge 1)$ in which 6 patients have been treated with $\le 1$
  instance of DLT, or dose level 0 if there were $\ge 2$ instances of
  DLT at dose level 1.''}
\end{quote}
\caption{\label{korn_protocol}Natural-language description of the
  \mbox{3 + 3} dose-escalation protocol, from \cite{korn_comparison_1994}.  Our
  specification implements the ``more common requirement'' described
  here.}
\end{figure}

\begin{figure}[h!]
\begin{quote}
{\em ``In the traditional 3 + 3, phase I cancer trial design, a minimum of
  three participants are studied at each dose level. If none of these
  three participants experience a DLT, a subsequent three participants
  are enrolled onto the next highest dose level. If one of three
  participants at a dose level experiences a DLT, up to three more
  participants are enrolled. When a DLT is observed in at least two
  participants in a cohort of three to six, the MTD is [regarded as
    having been] exceeded and an additional three participants (up to a
  total of six) are treated at the next lower dose level. The MTD is
  defined as the dose level at which none or one of six participants
  (0\% to 17\%) experience a DLT, when at least two of three to six
  participants (33\% to 67\%) experience a DLT at the next highest
  dose.''}
\end{quote}
\caption{\label{skolnik_protocol}Natural-language description of the
  \mbox{3 + 3} dose-escalation protocol, from \cite{skolnik_shortening_2008}.
  The ``more common'' variant of \cite{korn_comparison_1994} appears
  to be taken for granted here.}
\end{figure}

\section{Indeterminacies in dose escalation}\label{sec:indet}

Before we can reconstruct the aforementioned rules, we must appreciate
certain essential indeterminacies inherent to dose escalation.

Whereas a dose-escalation decision determines the dose at which the
next cohort of trial participants will enroll, it does not {\em fully}
determine the ensuing trial state.  This is because there is a nonzero
probability of DLT in any participant who receives a nonzero dose.
Thus, the toxicity tally in a cohort of size \verb"C" could turn out
as any of the \verb"C+1" possibilities \verb"[C/C,...,0/C]":

\begin{verbatim}
enroll_tallies(C, Qs) :-
    n_countdown(C, Ts),
    maplist(\T^(=(T/C)), Ts, Qs).

n_countdown(N, [N|Ns]) :- #N #> 0, #Nminus1 #= N - 1,
                          n_countdown(Nminus1, Ns).
n_countdown(0, [0]).
\end{verbatim}
Moreover, under an important generalization to be explored in
Section~\ref{sec:variations}, the actual number of patients who will
be available to enroll after a dose-escalation decision may itself be
indeterminate, drawn from a list \verb"[C|Cs]" of length $\ge 2$:
\begin{verbatim}
enrollments_tallies([C|Cs], Qs) :-
    enroll_tallies(C, Q0s),
    append(Q0s, Q1s, Qs),
    enrollments_tallies(Cs, Q1s).
enrollments_tallies([], []).
\end{verbatim}
Thus, the decision to enroll a new cohort into a dose level that
already has a \verb"T0/N0" toxicity tally gives rise to multiple
possible outcome \verb"Tallies":
\begin{verbatim}
tally0_tallies(T0/N0, Tallies) :-
    valid_tally(T0/N0),
    allowed_cohort_sizes(Cs),
    enrollments_tallies(Cs, Qs),
    maplist(\Qnew^Q^(Qnew=Tnew/Nnew,
                     #T #= T0 + Tnew,
                     #N #= N0 + Nnew, Q=T/N),
            Qs, Tallies).
\end{verbatim}

\section{Regret-constrained specification of $\mathbf{3 + 3}$ designs}\label{sec:rc}

We can now introduce what we call {\em regret-constrained} protocol
specifications, which ground dose escalation in specified events
deemed to induce clinically meaningful {\em decisional regret},
thereby constraining what dose-escalation decisions are permissible in
the course of a trial.  This renders explicit a linkage to underlying
aims and clinical motivations left implicit by customary treatments of
dose-escalation via painstaking procedural tabulations.  We will
exhibit a Prolog program that, on the basis of a few suitably calibrated yet
clinically intuitive regret-constraints, behaves consistently with the natural-language \mbox{3 + 3}
design rules quoted above.

To this end, we state that we \verb"regret" a \verb"Decision" when it
yields some undesirable \verb"TallyHistory", which for the purpose of
modeling the \mbox{3 + 3} design needs to take into account only the
resulting tally \verb"T/N" and the immediately preceding tally
\verb"T0/N0":

\begin{verbatim}
state0_decision_histories(S0, E, Hs) :-
    (   E = esc,
        S0 = [T0/N0|_] - [Told/Nold|_]
    ;   E = sta,
        S0 = [T0/N0|_] - _,
        Told = T0, Nold = N0
    ;   E = des,
        S0 = [T0/N0, Told/Nold | _] - _
    ),
    tally0_tallies(Told/Nold, Qs),
    maplist(\Q^(=([Q, T0/N0])), Qs, Hs).
\end{verbatim}

\subsection{Specific, clinically-motivated regrets}

It is always possible that, after an {\tt esc} decision, all three
patients in the new cohort will experience DLTs.  Although such an
{\em outcome} is of course regrettable, investigators need not regret
{\em their decision} to escalate if they feel it was justified by a
low enough toxicity observed at the previous dose.  The judgment as to
what constitutes {\em sufficiently low toxicity to justify escalation}
will generally depend on clinical context.  The \mbox{3 + 3} design
embeds a judgment equating this condition with {\tt N0}~$\ge 3$ and
{\tt T0/N0}~$\le 1/6 \in \mathbb{Q}$.  Accordingly, we regret
escalation precisely when this justification is absent:
\begin{verbatim}
decision_q0_q_regret(esc, T0/N0, _, #\ ( #N0 #>= 3 #/\ #T0 * 6 #=< N0 )).
\end{verbatim}
The literature makes equally clear that overly cautious
dose-escalation risks exposing too many participants to {\em
  subtherapeutic dosing} \cite{simon_accelerated_1997}.  Thus,
decisional regret can also serve to constrain {\tt des} decisions.
The \mbox{3 + 3} design is consistent with a judgment regretting
de-escalating from a moderately toxic tally $\{\mathtt{T0}\le 1,
\mathtt{N0}\ge 3\}$ upon appreciating very low net toxicity {\tt T/N}
$< 1/6$ at the new lower dose:
\begin{verbatim}
decision_q0_q_regret(des, T0/N0, T/N,
                     ( #T0 #=< 1 #/\ #N0 #>= 3 ) #/\
                     ( #N #> 0 #/\ #T * 6 #< #N )
                    ).
\end{verbatim}
The literature also exhibits ample evidence of interest in {\em net}
properties of dose-escalation trials {\em ex post facto}
\cite{italiano_treatment_2008}.  A \mbox{3 + 3} trial run without
protocol violations can never record more than 4 DLTs at any single
dose level.  We express this fact by positing a regret for {\em any}
decision that results in $\ge 5$ toxicities at any dose level,
regardless of preceding tally history:
\begin{verbatim}
decision_q0_q_regret(esc, _, T/_, #T #>= 5).
decision_q0_q_regret(sta, _, T/_, #T #>= 5).
decision_q0_q_regret(des, _, T/_, #T #>= 5).
\end{verbatim}
Prolog has allowed us to state these regrets in a very general form:
We regret {\em any} situation that satisfies the given constraints,
even if nothing else is known about it. Indeed, we capture an {\em
  infinite} set of concrete tallies with each of these clauses of
\verb"decision_q0_q_regret/4".

\subsection{Reification of regret}

To achieve efficiency while retaining the desired generality of our
code with the constructs from~\cite{DBLP:journals/corr/NeumerkelK16},
we {\em reify} these declared regrets in a predicate,
\verb"regret_t/3".  We ensure the \verb"true" and \verb"false"
branches of \verb"regret_t/3" are constructed without copy-and-paste error by
employing the \verb"term_expansion/2" mechanism to generate the
\verb"false" branch of \verb"regret_t/3" as the constructive negation
of all \verb"true" clauses, using the reification mechanism of
\verb"library(clpz)":
\begin{verbatim}
% Generate e_vars_disjunction _at compile time_.
:- dynamic(e_vars_disjunction/3).
term_expansion(generate_clauses, Clauses) :-
    findall(e_vars_disjunction(E, Vars, Disjunction),
            (   member(E, [esc,sta,des]),
                findall(Vs-RC,
                        (   decision_q0_q_regret(E, T0/N0, T/N, RC),
                            Vs = [N0,N,T0,T]
                        ), VsRCs),
                pairs_keys_values(VsRCs, Vs, RCs),
                maplist(=(Vars), Vs),
                foldl(\X^Disj0^Disj^(Disj=(Disj0#\/X)), RCs, 0#=1, Disjunction)
            ),
            Clauses).
generate_clauses.

regret_t(E, H, Truth)  :-
    e_h_disjunction(E, H, Disjunction),
    Disjunction #<==> #B,
    b_t(B, Truth).

b_t(0, false).
b_t(1, true).

e_h_disjunction(E, H, Disjunction) :-
    H = [T/N, T0/N0],
    e_vars_disjunction(E, [N0,N,T0,T], Disjunction).
\end{verbatim}

The \verb"e_vars_disjunction/3" clauses thus generated are:
\begin{verbatim}
?- listing(e_vars_disjunction/3).
 e_vars_disjunction(esc,[A,B,C,D],0#=1#\/ #\ (#A#>=3#/\ #C*6#=<A)#\/ #D#>=5).
 e_vars_disjunction(sta,[A,B,C,D],0#=1#\/ #D#>=5).
 e_vars_disjunction(des,[A,B,C,D],
                  0#=1#\/ #C#=<1#/\ #A#>=3#/\(#B#>0#/\ #D*6#< #B)#\/ #D#>=5).
    true.
\end{verbatim}

\subsection{Anticipatory regret}

Whereas \verb"regret_t/3" {\em looks backwards} to already-realized
toxicity tallies, it is also possible to {\em anticipate} all possible
\verb"Regrets":
\begin{verbatim}
state0_decision_regrets(S0, E, Regrets) :- 
    state0_decision_histories(S0, E, Hs),
    tfilter(regret_t(E), Hs, Regrets).
\end{verbatim}
A dose-escalation decision is \verb"regrettable" iff the list of
possible \verb"Regrets" is nonempty:
\begin{verbatim}
state0_decision_regrettable_t(S0, E, false) :- 
    state0_decision_regrets(S0, E, []).
state0_decision_regrettable_t(S0, E, true) :- 
    state0_decision_regrets(S0, E, [_|_]).
\end{verbatim}

By conducting dose escalation to {\em avert} anticipated regret,
resolving degenerate choices by therapeutically motivated preference
relations ${\tt esc} \succ {\tt sta} \succ {\tt des}$ that favor
exploration of higher doses, we obtain an effective protocol
specification.
\begin{verbatim}
state0_nextdecision(S0, E) :-
    if_((   state0_decision_infeasible_t(S0, esc)
        ;   state0_decision_regrettable_t(S0, esc)),
        if_((   state0_decision_infeasible_t(S0, sta)
            ;   state0_decision_regrettable_t(S0, sta)),
            if_((   state0_decision_infeasible_t(S0, des)
                ;   state0_decision_regrettable_t(S0, des)),
                E = stop,
                E = des),
            E = sta),
        E = esc).
\end{verbatim}

Finally, we obtain all possible trial paths using a DCG
\texttt{path//1} which describes all paths of the trial from the
current state~\texttt{S0}:

\begin{verbatim}
path(S0) --> { state0_nextdecision(S0, E),
               state0_decision_state(S0, E, S) },
             [E, S],
             path(S).
path(S0) --> { state0_nextdecision(S0, stop),
               stopstate_rec(S0, Rec) },
             [stop, recommend_dose(Rec)].
path(recommend_dose(_)) --> [].
\end{verbatim}
When none of the dose-escalation decisions {\tt [esc,sta,des]} remains
feasible, the trial {\tt stop}s, yielding a {\em dose recommendation}.
The relation between the final state and this recommendation is
specified by {\tt stopstate\_rec/2}:
\begin{verbatim}
stopstate_rec([]-_, 0).
stopstate_rec([T/N|TNs]-_, Rec) :- 
        (   #T * 6 #> #N, length(TNs, Rec)
        ;   #T * 6 #=< #N, length([_|TNs], Rec)
        ).
\end{verbatim}

\section{Algebraic properties of our formulation}

With the lone exception of \verb"once/1", which is guaranteed to be
used in a safe way as explained in Section~\ref{sec:mechanics}, the resulting program consists
exclusively of \textit{monotonic} Prolog constructs. Our program thus
\textit{by construction} provides algebraic properties that we
consider essential in this safety-critical application area:

\begin{itemize}
\item If an answer says that no (further) solutions exist to a query,
  then there are truly no (further) solutions, no matter which
  additional constraints we add. In other words, our program cannot be
  ``tricked'' into yielding solutions that other queries deny.
\item The predicates of \texttt{library(debug)} can be used for
  \textit{declarative debugging} of the program itself.
\item Every query either shows \textit{all} possible answers, or yields
  an \textit{instantiation error} if the query is not sufficiently
  instantiated. 
\item Reordering any clauses or goals in the program can
  affect termination or the existence of \textit{instantiation errors},
  but it does not change the set of described solutions.
\end{itemize}

In addition, the Scryer Prolog toplevel always shows all pending
constraints and can therefore be used like a theorem prover: When a
query succeeds unconditionally, a solution is guaranteed to exist.

These properties ensure that our program constitutes a declarative
specification that can be queried, run and reasoned about in
different~ways, and yields only correct results in all possible usage
modes. The only possible remaining source of logic~errors is a mistake
in the protocol formulation itself.

\section{Relation of our specification to natural-language descriptions}

From this program, we can elicit concretely much of the meaning of the
text in Figure~\ref{korn_protocol}, which we exhibit for the case of
${\tt D} = 3$ doses:
\begin{itemize}
\item {\em The dose levels are fixed in advance, with the first patients treated at dose level 1.}
  \item[]
\begin{verbatim}
?- setof(E^S1, Etc^(phrase(path([0/0]-[0/0,0/0]), [E, S1 | Etc])),
         Starts).
   Starts = [sta^([0/3]-[0/0,0/0]),sta^([1/3]-[0/0,0/0]),
             sta^([2/3]-[0/0,0/0]),sta^([3/3]-[0/0,0/0])].
\end{verbatim}

\medskip\item {\em If 0 out of 3 patients experiences DLT, one proceeds to the next higher level with a cohort of 3 patients.}
\item[]
\begin{verbatim}
?- setof(E, Path^Ls^H^Hs^(phrase(path([0/0]-[0/0,0/0]), Path),
             phrase((..., [[0/3|Ls]-[H|Hs]], [E], ...), Path)), Es).
   Es = [esc].
\end{verbatim}

\medskip\item {\em If 1 out of 3 patients experiences DLT, treat an additional 3 patients at the same dose level.}
\item[]
\begin{verbatim}
?- setof(E, Path^Ls^H^Hs^(phrase(path([0/0]-[0/0,0/0]), Path),
             phrase((..., [[1/3|Ls]-[H|Hs]], [E], ...), Path)), Es).
   Es = [sta].
\end{verbatim}

\medskip\item {\em If 1 out of 6 patients experiences DLT at a level, the dose escalates for the next cohort of patients.}
\item[]
\begin{verbatim}
?- setof(E, Path^Ls^H^Hs^(phrase(path([0/0]-[0/0,0/0]), Path),
                          H = 0/0,
                phrase((...,[[1/6|Ls]-[H|Hs]],[E],...), Path)), Es).
   Es = [esc].
\end{verbatim}
\end{itemize}
Notably, eliciting the intended meaning of the text in the last case
required positing that the next-higher dose \verb"H" had not yet been
enrolled. Without this proviso, \verb"stop" would also have been
permissible.

From the decision cascade in the body of \verb"state0_nextdecision/2"
it is evident that our formulation yields a unique decision in any
concrete situation (with ground \verb"S0"); we therefore consider it a complete specification.
The Appendix includes a complete enumeration of all 46 paths for the
${\tt D} = 2$ case, easily checked vis-à-vis the text of
Figure~\ref{korn_protocol}.

\section{Verification of safety properties}\label{sec:safety}

The text of Figure~\ref{korn_protocol} trends from language that is
initially imperative, toward declarative statements near the end. The
latter reflect clinical investigators' intense interest in asserting
{\em safety properties} for their trials and (by extension) for the
conclusions drawn from them. Informally, a safety property states that
{\em bad things do not happen}. For example, the ``more common
requirement''
\begin{quote}
  \begin{center}
    {\em to have 6 patients treated at the MTD (if it is higher than level 0)}
  \end{center}
\end{quote}
expresses a concern to avoid recommending a dose on the strength of
too little experience.

Our executable specification exhibits safety properties expressed in
Figure~\ref{korn_protocol} through statements about `the MTD'.
For example,
\begin{itemize}
\medskip\item {\em If $\ge 2$ out of 6 patients experience DLT, or $\ge 2$ out
  of 3 patients experience DLT in the initial cohort treated at a dose
  level, then one has exceeded the MTD.}
\item[]
\begin{verbatim}
recommends_dose_exceeding_mtd             :- 
   D in 1..8, indomain(D),       % For trials of up to D=8 doses
   InitD = [Q]-Qs, length([Q|Qs], D), % .. that start from the lowest dose
   maplist(=(0/0), [Q|Qs]),      % .. with no prior toxicity information,
   phrase(path(InitD), Path),    % does any Path exist
   phrase((..., [Ls-_], ...,     % .. on which a state Ls-_ appears,
           [recommend_dose(Rec)] % .. such that the recommended dose Rec
          ), Path),
   length(Ls,X), Rec #>= X,      % .. was at least the current dose X,
   Ls = [T/_|_], #T #> 1.        % .. yet X `exceeded MTD' per protocol?
%?- time(recommends_dose_exceeding_mtd).
%@    % CPU time: 574.029s, 2_701_969_903 inferences
%@    false. % This safety property is verified for trials of up to 8 doses.
\end{verbatim}
\end{itemize}
Note that the featured statement effectively asserts a property
of the desirable {\em dose recommendation} (which it refers to as `the
MTD'), namely that it should be less than any dose at which $\ge 2$ toxicities
have been tallied.

\section{Verification of liveness properties}

Although clinical investigators' keen attention to safety is more
readily apparent in the literature, any sensible dose-escalation
design must also guarantee {\em liveness} properties.  Informally,
liveness properties state that {\em good things do happen}.  One such
property applicable to these trials is that they {\em yield a
  recommendation, and then immediately stop}:
\begin{verbatim}
trial_fails_to_conclude_with_unique_rec :-
   D in 1..8, indomain(D), length([Q|Qs], D), maplist(=(0/0), [Q|Qs]),
   phrase(path([Q]-Qs), Path),
   (   phrase((..., [recommend_dose(Rec)], [_], ...), Path)
   ;   \+ phrase((..., [recommend_dose(Rec)]), Path)
   ).
%?- time(trial_fails_to_conclude_with_unique_rec).
%@    % CPU time: 560.946s, 2_675_988_416 inferences
%@    false. % All trials of up to 8 doses yield a recommendation, then stop.
\end{verbatim}


\section{Generality and versatility of the specification}

The DCG introduced in Section~\ref{sec:safety} is remarkably general
and versatile. It is an {\em executable specification} of the
sequences of events which may occur in the trial, in the sense that it
states \textit{what~holds}, and it can be used in multiple
\textit{modes}. This generality allows us to answer a variety of
questions with comparatively simple queries. For example, in one
specific mode, we can use our specification as an `expert system' that
automatically determines what the next decision should be at a point
where we have reached the highest dose of a ${\tt D} = 3$ trial,
having recorded a 0/3 tally at each dose:
\begin{verbatim}
?- setof(E, Path^(phrase(path([0/0]-[0/0,0/0]), Path),
                  phrase((...,[[0/3,0/3,0/3]-[], E], ...), Path)), Es).
   Es = [sta].
\end{verbatim}
We can as easily adopt a retrospective point of view, to ask what
dose-escalation decisions {\em could have preceded} this state:
\begin{verbatim}
?- setof(E0, Path^(phrase(path([0/0]-[0/0,0/0]), Path),
                   phrase((...,[E0, [0/3,0/3,0/3]-[]],...), Path)), E0s).
   E0s = [esc].
\end{verbatim}
We are also able to look far ahead, anticipating all recommendations
that remain possible at any time in an ongoing trial.  Having recorded
tallies of 0/3, 0/3 and 2/6 at dose levels 1, 2 and 3 of a ${\tt D} = 3$
trial, for example:
\begin{verbatim}
?- setof(Rec, Path^(phrase(path([0/0]-[0/0,0/0]), Path),
                    phrase((..., [[2/6,0/3,0/3]-[]],
                            ..., [recommend_dose(Rec)]), Path)), Recs).
   Recs = [0,1,2]. % Only dose level 3 has been ruled out so far.
\end{verbatim}
Finally, the DCG allows complete enumeration of \textit{all admissible
  trials}:
\begin{verbatim}
?- J+\time((D = 8, length([Q|Qs], D), maplist(=(0/0), [Q|Qs]),
            setof(Path, phrase(path([Q]-Qs), Path), Paths),
            length(Paths, J))).
%@    % CPU time: 337.638s, 1_573_175_565 inferences
%@    J = 16138.
\end{verbatim}

This capability can confer substantial benefits even on analyses of a
statistical nature. For example, \cite{norris_what_2020} employs such
enumeration to derive statistical characterizations of \mbox{3 + 3}
trials, exempt from Monte Carlo error. The generality and versatility
of our Prolog~code enables more usage modes than a procedural
implementation can provide. We regard the run-times of our program in
all these modes as commensurate to the importance of reliable
guarantees in this domain, and therefore do not pursue any further
performance considerations in the present paper.

\section{Flexibly accommodating protocol variations}\label{sec:variations}

As shown by the expansive tabulations in
\cite{skolnik_shortening_2008} and \cite{frankel_model_2020}, when
working within the tradition of informal protocol specifications, much
labor is needed to adapt the typically lumpy cohorts of
dose-escalation designs to trial enrollment occurring as a rather
fluid {\em arrival process} in real time.  Our RC specification
framework, by contrast, shows promise for being readily adaptable to
this challenge.

If we recompile our program with an expanded set of possible cohort sizes,
\begin{verbatim}
allowed_cohort_sizes([3,2,1]).
\end{verbatim}
then, as the following query shows, this added flexibility renders the
protocol adequate to admit paths in which participants enroll either
singly or in pairs, rather than waiting for arrival of a third
participant to complete the traditional cohort of 3.  Such
just-in-time or `rolling' \cite{skolnik_shortening_2008} enrollment
may enable de-escalation decisions to occur sooner, avoiding dosing
patients at levels that are no longer viable dose
recommendations. This improves safety, and also accelerates progress
of the trial.
\begin{verbatim}
?- phrase(path([0/3,0/3,0/3]-[]), [sta, [2/5,0/3,0/3]-[],
                                   des, [0/6,0/3]-[2/5],
                                   stop, recommend_dose(2)]).
   true % The above-described path is admissible.
;  ... .
\end{verbatim}

\section{Conclusion}

Possibly the most essential aspect of our program's {\em
  declarativeness} is its generality.  This quality has allowed us
to examine our program in numerous ways to probe comprehensively for
various forms of error.
We can investigate individual predicates with general queries that
expose their full meaning {\em as implemented}, to check for departures
from {\em intended} meaning.
Yet the very same program allows us to recapitulate the best
available natural-language description of the \mbox{3 + 3} protocol,
and indeed in so doing to recognize deficiencies in that description
and {\em correct} them.  Not only does this reduce the likelihood that
we have failed to model `the true' \mbox{3 + 3} protocol, but it gives
us a claim to have definitively superseded the extant natural-language
descriptions.

The {\em internal} consistency of our program, on the other hand,
derives from its very construction using a pure monotonic subset of Prolog.
This enables us to trust \verb"false"
answers from queries that search for violations of safety and liveness
properties, rendering such answers {\em guarantees}.




\begin{credits}
\subsubsection{\ackname}
We thank Ulrich Neumerkel and several anonymous reviewers for their valuable comments.

\subsubsection{\discintname}
The authors have no competing interests to declare that are
relevant to the content of this article. This work received no external funding.
\end{credits}
%
%
%
%
\newpage
\bibliography{FLOPS2024,DPLB}

\newpage
\section*{Appendix}

A complete listing of all 46 paths of the \mbox{3 + 3} design for the
${\tt D} = 2$ case.  These may be checked one-by-one against the text
of Figure~\ref{korn_protocol}.

\scriptsize
\begin{verbatim}
:- use_module(library(format)).
?- J+\(setof(Path, (phrase(path([0/0]-[0/0]), Path)), Paths)
      , maplist(portray_clause, Paths), length(Paths, J)).
[sta,[0/3]-[0/0],esc,[0/3,0/3]-[],sta,[0/6,0/3]-[],stop,recommend_dose(2)].
[sta,[0/3]-[0/0],esc,[0/3,0/3]-[],sta,[1/6,0/3]-[],stop,recommend_dose(2)].
[sta,[0/3]-[0/0],esc,[0/3,0/3]-[],sta,[2/6,0/3]-[],des,[0/6]-[2/6],stop,recommend_dose(1)].
[sta,[0/3]-[0/0],esc,[0/3,0/3]-[],sta,[2/6,0/3]-[],des,[1/6]-[2/6],stop,recommend_dose(1)].
[sta,[0/3]-[0/0],esc,[0/3,0/3]-[],sta,[2/6,0/3]-[],des,[2/6]-[2/6],stop,recommend_dose(0)].
[sta,[0/3]-[0/0],esc,[0/3,0/3]-[],sta,[2/6,0/3]-[],des,[3/6]-[2/6],stop,recommend_dose(0)].
[sta,[0/3]-[0/0],esc,[0/3,0/3]-[],sta,[3/6,0/3]-[],des,[0/6]-[3/6],stop,recommend_dose(1)].
[sta,[0/3]-[0/0],esc,[0/3,0/3]-[],sta,[3/6,0/3]-[],des,[1/6]-[3/6],stop,recommend_dose(1)].
[sta,[0/3]-[0/0],esc,[0/3,0/3]-[],sta,[3/6,0/3]-[],des,[2/6]-[3/6],stop,recommend_dose(0)].
[sta,[0/3]-[0/0],esc,[0/3,0/3]-[],sta,[3/6,0/3]-[],des,[3/6]-[3/6],stop,recommend_dose(0)].
[sta,[0/3]-[0/0],esc,[1/3,0/3]-[],sta,[1/6,0/3]-[],stop,recommend_dose(2)].
[sta,[0/3]-[0/0],esc,[1/3,0/3]-[],sta,[2/6,0/3]-[],des,[0/6]-[2/6],stop,recommend_dose(1)].
[sta,[0/3]-[0/0],esc,[1/3,0/3]-[],sta,[2/6,0/3]-[],des,[1/6]-[2/6],stop,recommend_dose(1)].
[sta,[0/3]-[0/0],esc,[1/3,0/3]-[],sta,[2/6,0/3]-[],des,[2/6]-[2/6],stop,recommend_dose(0)].
[sta,[0/3]-[0/0],esc,[1/3,0/3]-[],sta,[2/6,0/3]-[],des,[3/6]-[2/6],stop,recommend_dose(0)].
[sta,[0/3]-[0/0],esc,[1/3,0/3]-[],sta,[3/6,0/3]-[],des,[0/6]-[3/6],stop,recommend_dose(1)].
[sta,[0/3]-[0/0],esc,[1/3,0/3]-[],sta,[3/6,0/3]-[],des,[1/6]-[3/6],stop,recommend_dose(1)].
[sta,[0/3]-[0/0],esc,[1/3,0/3]-[],sta,[3/6,0/3]-[],des,[2/6]-[3/6],stop,recommend_dose(0)].
[sta,[0/3]-[0/0],esc,[1/3,0/3]-[],sta,[3/6,0/3]-[],des,[3/6]-[3/6],stop,recommend_dose(0)].
[sta,[0/3]-[0/0],esc,[1/3,0/3]-[],sta,[4/6,0/3]-[],des,[0/6]-[4/6],stop,recommend_dose(1)].
[sta,[0/3]-[0/0],esc,[1/3,0/3]-[],sta,[4/6,0/3]-[],des,[1/6]-[4/6],stop,recommend_dose(1)].
[sta,[0/3]-[0/0],esc,[1/3,0/3]-[],sta,[4/6,0/3]-[],des,[2/6]-[4/6],stop,recommend_dose(0)].
[sta,[0/3]-[0/0],esc,[1/3,0/3]-[],sta,[4/6,0/3]-[],des,[3/6]-[4/6],stop,recommend_dose(0)].
[sta,[0/3]-[0/0],esc,[2/3,0/3]-[],des,[0/6]-[2/3],stop,recommend_dose(1)].
[sta,[0/3]-[0/0],esc,[2/3,0/3]-[],des,[1/6]-[2/3],stop,recommend_dose(1)].
[sta,[0/3]-[0/0],esc,[2/3,0/3]-[],des,[2/6]-[2/3],stop,recommend_dose(0)].
[sta,[0/3]-[0/0],esc,[2/3,0/3]-[],des,[3/6]-[2/3],stop,recommend_dose(0)].
[sta,[0/3]-[0/0],esc,[3/3,0/3]-[],des,[0/6]-[3/3],stop,recommend_dose(1)].
[sta,[0/3]-[0/0],esc,[3/3,0/3]-[],des,[1/6]-[3/3],stop,recommend_dose(1)].
[sta,[0/3]-[0/0],esc,[3/3,0/3]-[],des,[2/6]-[3/3],stop,recommend_dose(0)].
[sta,[0/3]-[0/0],esc,[3/3,0/3]-[],des,[3/6]-[3/3],stop,recommend_dose(0)].
[sta,[1/3]-[0/0],sta,[1/6]-[0/0],esc,[0/3,1/6]-[],sta,[0/6,1/6]-[],stop,recommend_dose(2)].
[sta,[1/3]-[0/0],sta,[1/6]-[0/0],esc,[0/3,1/6]-[],sta,[1/6,1/6]-[],stop,recommend_dose(2)].
[sta,[1/3]-[0/0],sta,[1/6]-[0/0],esc,[0/3,1/6]-[],sta,[2/6,1/6]-[],stop,recommend_dose(1)].
[sta,[1/3]-[0/0],sta,[1/6]-[0/0],esc,[0/3,1/6]-[],sta,[3/6,1/6]-[],stop,recommend_dose(1)].
[sta,[1/3]-[0/0],sta,[1/6]-[0/0],esc,[1/3,1/6]-[],sta,[1/6,1/6]-[],stop,recommend_dose(2)].
[sta,[1/3]-[0/0],sta,[1/6]-[0/0],esc,[1/3,1/6]-[],sta,[2/6,1/6]-[],stop,recommend_dose(1)].
[sta,[1/3]-[0/0],sta,[1/6]-[0/0],esc,[1/3,1/6]-[],sta,[3/6,1/6]-[],stop,recommend_dose(1)].
[sta,[1/3]-[0/0],sta,[1/6]-[0/0],esc,[1/3,1/6]-[],sta,[4/6,1/6]-[],stop,recommend_dose(1)].
[sta,[1/3]-[0/0],sta,[1/6]-[0/0],esc,[2/3,1/6]-[],stop,recommend_dose(1)].
[sta,[1/3]-[0/0],sta,[1/6]-[0/0],esc,[3/3,1/6]-[],stop,recommend_dose(1)].
[sta,[1/3]-[0/0],sta,[2/6]-[0/0],stop,recommend_dose(0)].
[sta,[1/3]-[0/0],sta,[3/6]-[0/0],stop,recommend_dose(0)].
[sta,[1/3]-[0/0],sta,[4/6]-[0/0],stop,recommend_dose(0)].
[sta,[2/3]-[0/0],stop,recommend_dose(0)].
[sta,[3/3]-[0/0],stop,recommend_dose(0)].
   J = 46.
\end{verbatim}

\end{document}